\documentstyle[sprocl,epsfig,citesort]{article}

\def\be{\begin{equation}}
\def\ee{\end{equation}}
\def\bea{\begin{eqnarray}}
\def\eea{\end{eqnarray}}
\newcommand{\EPEM}{\mbox{e$^+$e$^-$}}
\newcommand{\GG}{\mbox{$\gamma\gamma$}}
\newcommand{\GE}{\mbox{$\gamma$e}}

\newcommand{\ggww}{\mbox{$\gamma\gamma\to \textrm{\rm W}^+\textrm{\rm W}^-\,$}}
\newcommand{\sigmaw}{\mbox{$\sigma_W\,$}}
\newcommand{\mww}{\mbox{$M_W^2\,$}}
\newcommand{\mw}{\mbox{$M_W\,$}}
\newcommand{\mz}{\mbox{$M_Z\,$}}
\newcommand{\gewnu}{\mbox{$\gamma e\to \textrm{\rm W}\nu\,$}}

\newcommand{\ggam}{\mbox{$\gamma\gamma\,$}}

\def\fbi{~{\rm fb}^{-1}}
\def\rts{\sqrt s}
\def\gev{\,{\rm GeV}}

\def\br{BR}
\def\hsm{h_{\rm SM}}
\def\gam{\gamma}
\def\anti{\overline}
\def\gamhsm{\Gamma_{\hsm}^{\rm tot}}
\def\epem{e^+e^-}
\def\mupmum{\mu^+\mu^-}
\def\thdm{2HDM}
\def\mhsm{m_{\hsm}}
\def\wstar{W^{\star}}
\def\gamgam{\gam\gam}
\def\lsim{\mathrel{\raise.3ex\hbox{$<$\kern-.75em\lower1ex\hbox{$\sim$}}}}
\def\gsim{\mathrel{\raise.3ex\hbox{$>$\kern-.75em\lower1ex\hbox{$\sim$}}}}
\def\textrm{\rm}
\def\ie{{\it i.e.}}
\def\mt{m_t}
\def\hl{h^0}
\def\mhl{m_{\hl}}
\def\g{\gamma}
\def\l{\lambda}
\def\s{\sigma}

\begin{document}

\title{Physics at the Photon Linear Collider
\footnote{Talk presented at the Photon'97 Conference,
Egmond aan Zee, the Netherlands, May 10-15, 1997.}}

\author{ G.~JIKIA\footnote{Alexander von Humboldt Fellow;
	e-mail: jikia@phyv4.physik.uni-freiburg.de}}

\address{
Albert--Ludwigs--Universit\"{a}t Freiburg, Fakult\"{a}t f\"{u}r Physik\\
Hermann--Herder Str.3, D-79104 Freiburg, Germany \\
and \\
Institute for High Energy Physics, Protvino\\
Moscow Region 142284, Russian Federation}

%%%%%%%%%%%%%%%%%%%%%%%%%%%%%%%%%%%%%%%%%%%%%%%%%%%%%%%%%%%%%%
% You may repeat \author \address as often as necessary      %
%%%%%%%%%%%%%%%%%%%%%%%%%%%%%%%%%%%%%%%%%%%%%%%%%%%%%%%%%%%%%%

\rightline{Freiburg--THEP 97/11}
\rightline{hep-ph/9706508}
\rightline{June 1997}

\vspace*{1cm}

\maketitle\abstracts{The physics prospects of the high energy Photon
Linear Collider are reviewed, emphasizing its potential to study the
symmetry breaking sector, including Higgs searches and precision
anomalous $W$ couplings measurements.}

\section{Introduction}

Using the process of Compton backscattering of laser light off
electron beams from the linear collider one can obtain \GG\ and \GE\
colliding beams with an energy and luminosity comparable to that in
\EPEM\ collisions \cite{TEL97}. The expected physics at the Photon
Linear Collider (PLC) is very rich and complementary to that in \EPEM\
collisions. Since there exist several excellent extensive reviews on
the subject \cite{BRODHAW,BAIL94,BRODBER,CHANBER,GINSH,Belanger95}
only the issues concerning the electroweak physics based on recent
progress that has been achieved since Photon'95 Conference in
Sheffield are summarized here.

\section{ Higgs boson physics}

Discovery and study of Higgs boson(s) will be of primary importance at
future $pp$ and linear $e^+e^-$ and \GG\ colliders. The survey of the
Higgs physics opportunities of PLC is simultaneously a very good
example showing how the complete phenomenological portrait is obtained
only by combining the complementary information available from these
distinct types of machines.

\subsection{Measurements of the Higgs boson couplings}

The most fundamental properties of the Higgs boson are its mass, its
total width and its partial widths.  Ideally, one would wish to
determine, {\it in a model-independent way}, all of the tree-level and
one-loop couplings of the $h^0$, its spin, parity, and $CP$ nature, and
its total width.  The total Higgs width, while certainly important in
its own right, becomes even more so since it is required in order to
compute many important partial widths, which provide the most
direct means of verifying that the observed Higgs boson is or is not
the $\hsm$. While branching ratios, being the ratio of a partial width to
the total width can not be unambiguously interpreted,  any deviations
of partial widths from SM predictions can be directly compared to
predictions of alternative models such as the MSSM, the Non-Minimal 
Supersymmetric Standard Model, or the
general two Higgs doublet model (\thdm) \cite{snow96}. 

\begin{figure}[htb]
\setlength{\unitlength}{1in}
\begin{picture}(4.7,2.15)
\put(-.05,-.1){\epsfig{file=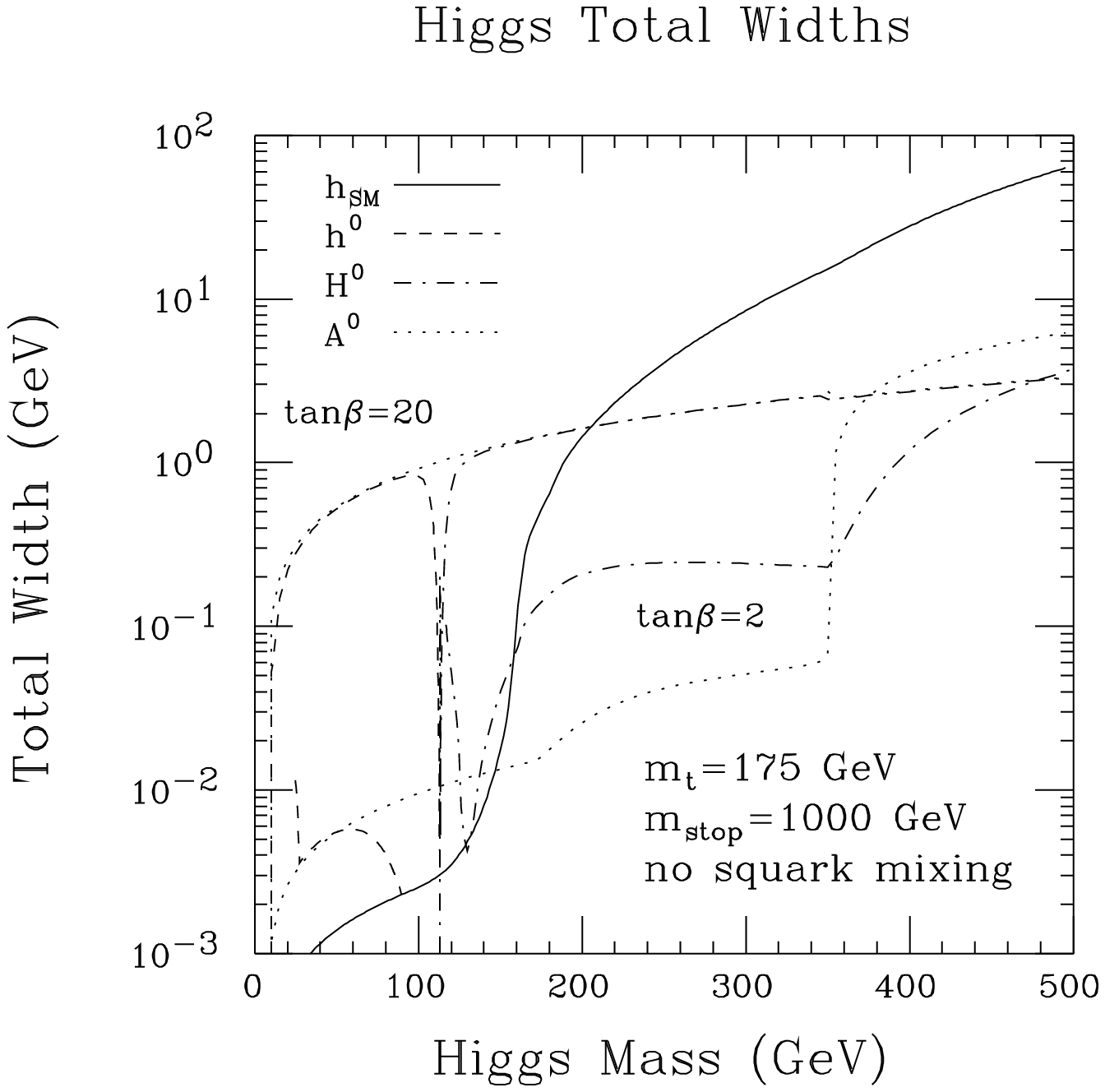,width=2.35in}}
\put(2.8,0.4){\epsfig{bbllx=40pt,bblly=195pt,bburx=426pt,bbury=574pt,file=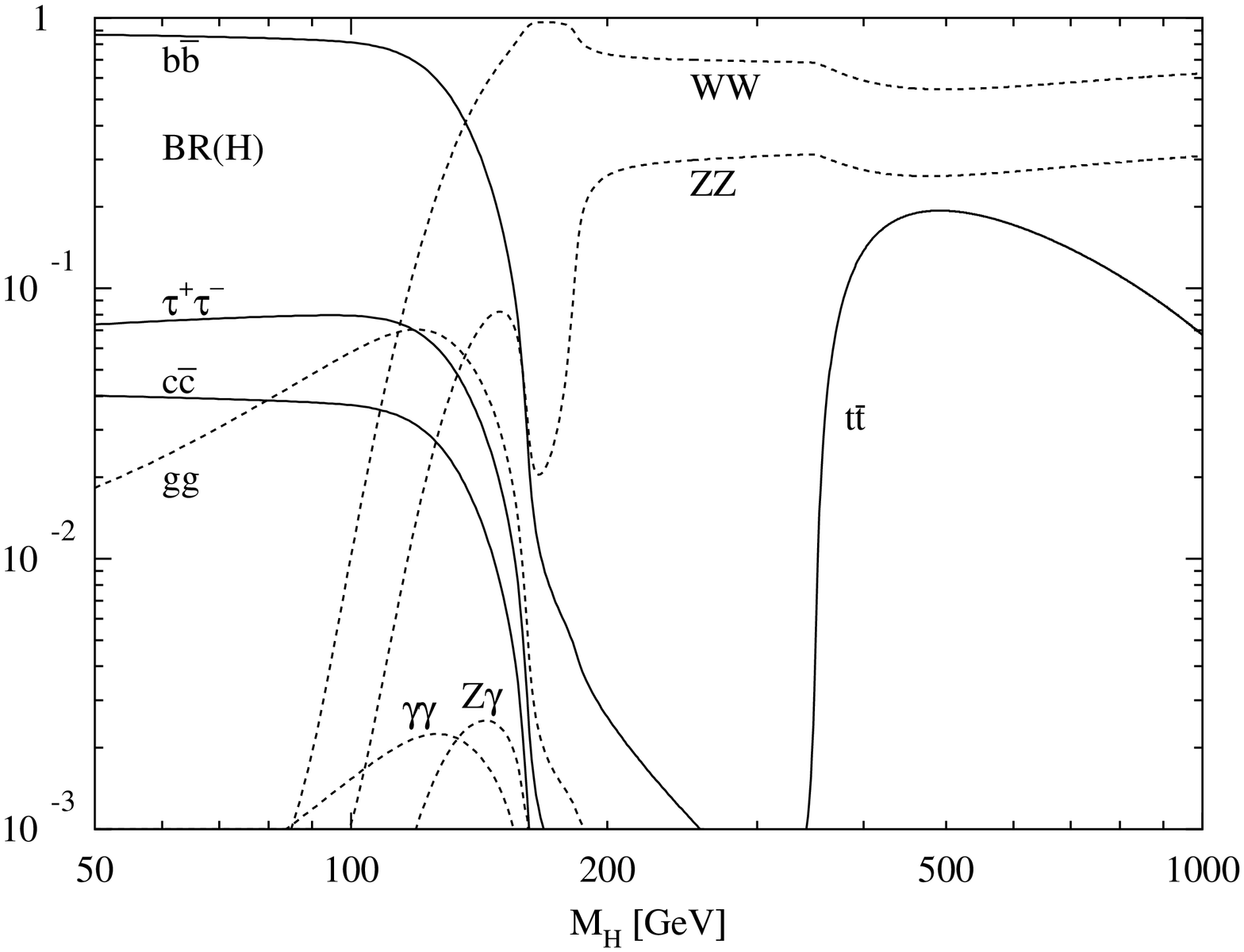,width=1.2in,height=1.62in,angle=270}}
\end{picture}
\caption[]{Total width versus mass of the SM and MSSM Higgs bosons for
$\mt=175\gev$ \cite{snow96} and the main branching ratios $BR(\hsm)$ of
the Standard Model Higgs decay channels \cite{hdecay}.}
\label{fig:width}
\end{figure}

The predicted width, $\gamhsm$, and branching ratios are plotted in
Fig.~\ref{fig:width} as a function of $\mhsm$. For $\mhsm\lsim 2\mw$,
$\gamhsm$ is too small to be reconstructed in the final state and only
indirect determination of $\gamhsm$ is possible at NLC and LHC using a
multiple step process; the best process depends upon
the Higgs mass.  In this respect \GG\ collider mode offers a unique
possibility \cite{BARKLOW,GunionHaber,BBC} to produce the Higgs boson
as an $s$-channel resonance decaying, for instance, into $b\bar b$:
\begin{equation}
\gamma\gamma\to h^0 \to b \bar b
\label{gghbb}
\end{equation}
and thereby measuring the rate for the Higgs boson production in \GG\
mode of the linear collider we can determine the value of the Higgs
two-photon width itself.  Assuming that 300-500~GeV linear collider
will first start operating in $e^+e^-$ mode, the mass of the $h^0$
will already be known from the Bjorken reaction $e^+e^-\to Z^* \to
Zh$, and the beam energy could be tuned so that the \GG\ luminosity
spectrum peaks at $m_h$.  The Higgs two-photon decay width is of
special interest by itself since it appears at the one-loop level.
Thus, any heavy charged particles which obtain their masses from
electroweak symmetry breaking can contribute in the loop.  Moreover,
for $\mhsm\lsim 130\gev$ (\ie\ in the MSSM $\mhl$ range), {\it the
only known procedure
\footnote{The other alternative is to employ FMC
$\mupmum$ collisions at $\rts\sim \mhsm$ and directly measure
$\gamhsm$ by scanning \cite{snow96}.}
 for determining the total Higgs width} $\Gamma^{tot}(h)$
is that based on the measurement of $\Gamma(h\to\gam\gam)$ in the
reaction (\ref{gghbb}) as described in Ref.~\cite{snow96}.

The following procedure could be used.  First one should measure the
cross section of the single Higgs production at PLC
\begin{equation}
\sigma(\GG\to h^0\to X) = \tau \frac{dL_{\GG}}{d\tau} 
\frac{8\pi^2}{m_h^3}
\Gamma(h^0\to\gamma\gamma)\cdot BR(h^0\to X)
(1+\lambda_1\lambda_2)
\label{siggghbb}
\end{equation}
and determine $\Gamma(h\to \gam\gam)\br(h\to b\anti b)$.  Here
the effective photon-photon luminosity $L_{\g\g}$ is introduced,
$\tau=m_h^2/s$.  Then
one can compute the two-photon width as a ratio
\be
\Gamma(h\to \gamma\gamma) = 
\frac{[\Gamma(h\to\gamma\gamma)BR(h\to b\bar b)]}{BR(h\to b\bar b)}.
\ee
The branching ratio $BR(h\to b\bar b)$ will also already be known
from $e^+e^-$ annihilation. Indeed, measuring $\sigma(\epem\to ZH)$
(in the missing mass mode) and $\sigma(\epem\to ZH)BR(h\to b\bar b)$
in $e^+e^-$ mode of the linear collider we can compute
\be
BR(h\to b\bar b) = \frac{[\sigma(\epem\to ZH)BR(h\to b\bar b)]}{
\sigma(\epem\to ZH)},
\ee
the error in the branching ratio is estimated at $\pm(5\div 10)\%$
\cite{snow96}.
Finally, one can compute the total Higgs boson width
\be
\Gamma^{tot}_h=\frac{\Gamma(h\to\gamma\gamma)}{BR(h\to\gamma\gamma)},
\ee
using the $\br(h\to\gam\gam)$ determination(s) at NLC and LHC
\cite{snow96}
\begin{eqnarray}
BR(h\to\gamma\gamma)& = &BR(h\to b\bar b)
\frac{[\sigma(\epem\to Zh)BR(h\to\gamma\gamma)]}
{[\sigma(\epem\to Zh)BR(h\to b\bar b)]}
\\
& = & BR(h\to b\bar b)
\frac{[\sigma(pp\to Wh)BR(h\to\gamma\gamma)]}
{[\sigma(pp\to Wh)BR(h\to b\bar b)]}
\nonumber
\end{eqnarray}
and compute {\it in a model-independent way} partial Higgs decay widths that
are directly related to fundamental couplings
\begin{equation}
\Gamma(h\to b\bar b) =\Gamma^{tot}_h BR(h\to b\bar b),\
\Gamma(h\to gg) =\Gamma^{tot}_h BR(h\to gg)\dots
\end{equation}
The observable cross section for the \GG\ Higgs signal in the gluon
fusion reaction at the LHC can depend quite strongly on the masses and
couplings of the superpartners and Higgs bosons, particularly if they
are not too heavy, and it varies from a few fb to more than 100~fb
over the parameter space of the MSSM, even in the scenario that
supersymmetry is not discovered at LEP2 \cite{KKMW}. Having measured
$BR(h\to gg)\cdot \Gamma(h\to\gamma\gamma)$ (with an error of order
$\pm 22\%$ at $m_{h_{SM}}=120$~GeV \cite{LHC}) and combining this
number with the value of the Higgs total and two-photon decay width,
measured in \GG\ and $e^+e^-$ experiments one can calculate the
two-gluon Higgs branching ratio and partial width.

\begin{figure}[htb]
\setlength{\unitlength}{1in}
\begin{picture}(4.7,2.05)
\put(1.3,-.2){\epsfig{file=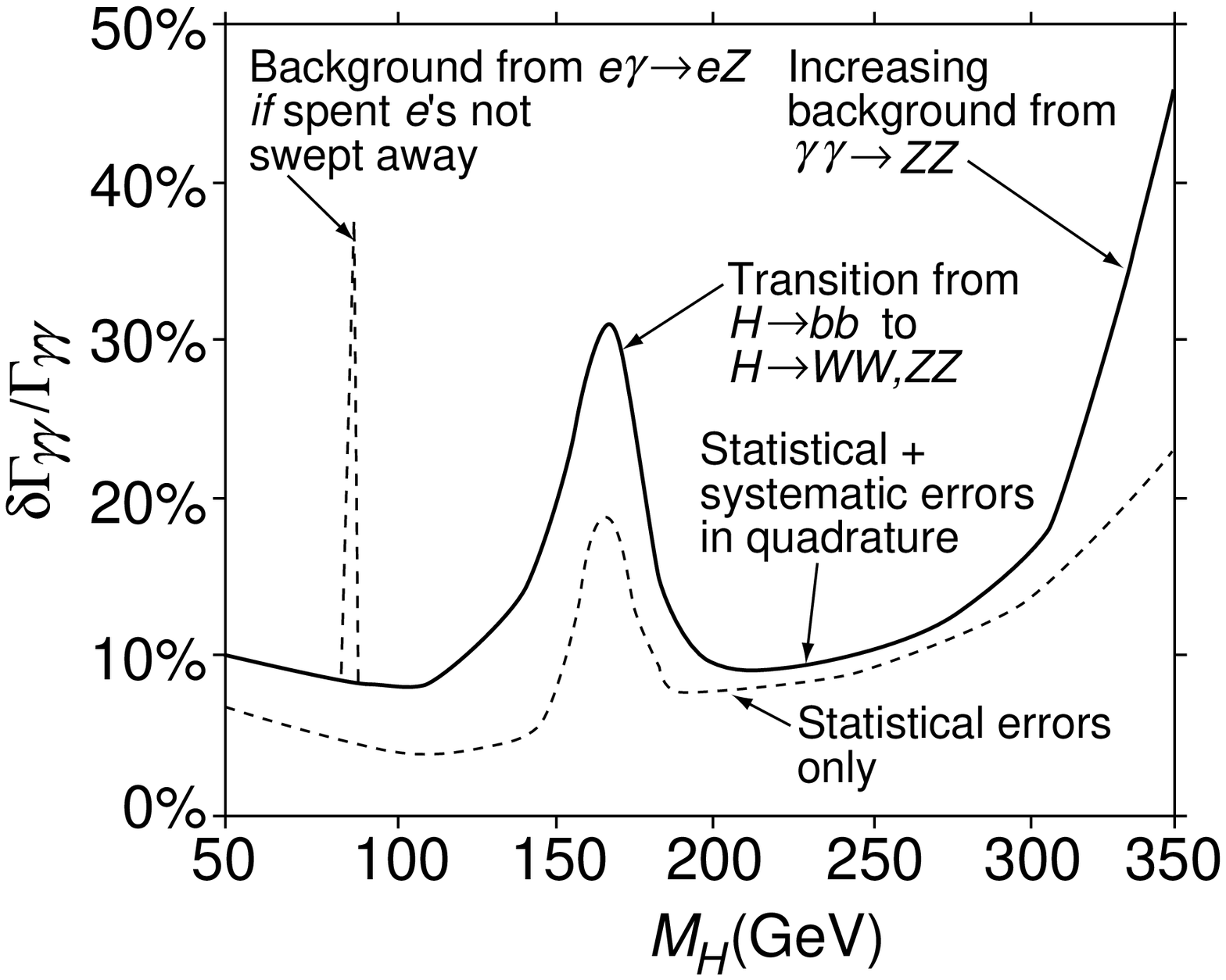,width=2.35in,height=2.35in}}
\end{picture}
\caption{Accuracy 
with which $\Gamma(\hsm\to \gam\gam)\br(\hsm\to b\anti b~{\rm or} WW,ZZ)$
can be measured at the PLC \protect\cite{snow96}.}
\label{fig:gamma-gamma}
\end{figure}

The main background to the $h^0$ production is the continuum
production of $b\bar b$ and $c\bar c$ pairs. In this respect, the
availability of high degree of photon beams circular polarization is
crucial, since for the equal photon helicities $(\pm\pm)$ that produce
spin-zero resonant states, the $\gamma\gamma\to q\bar q$ QED Born
cross section is suppressed by the factor $m_q^2/s$
\cite{BARKLOW,BBC,GunionHaber}. Another potentially dangerous
backgrounds originate from the resolved-photon processes
\cite{EGGHZ,BBB,JikiaTkabladze} in which a gluon from the photon
structure function produces $b\bar b$, $c\bar c$ pairs, and from the
continuum production of $b\bar b$ pairs accompanied by the radiation
of additional gluon \cite{Khhiggs,OhgakiTakahashi}, calculated taking
into account large QCD ${\cal O}(\alpha_s)$ radiative corrections
\cite{JikiaTkabladze}, which are not suppressed even for the equal
photon helicities. Virtual one-loop QCD corrections for $J_z=0$ were
found to be especially large due to the double logarithmic enhancement
factor, so that the corrections are comparable or even larger than the
Born contribution for the two-jet final topologies
\cite{JikiaTkabladze}. For small values of the cutoff $y_{cut}$,
separating two and three-jet events, two-jet cross section, calculated
to order $\alpha_s$, becomes even negative in the central region.
Recently leading QCD corrections for $J_z=0$ have been calculated at
the two-loop level \cite{FadinKhozeMartin}. The non-Sudakov form
factor in the double logarithmic approximation, including the two-loop
contribution \cite{FadinKhozeMartin}, is given by 
\be
\frac{\sigma_{2-loop}}{\sigma_{Born}}\Biggr|_{J_z=0} = 
1-2\frac{\alpha_s}{\pi}
\log^2(\frac{s}{m_b^2})+\frac{121}{108}\left(\frac{\alpha_s}{\pi}\right)^2
\log^4(\frac{s}{m_b^2}).  
\ee 
The account of two-loop contribution makes cross section to be
positive and the authors of Ref. \cite{FadinKhozeMartin} argue that
the higher order contributions are not so anomalously large.  Anyway,
these detailed studies
\cite{JikiaTkabladze,Khhiggs,BBB,OhgakiTakahashi,FadinKhozeMartin,snow96}
have shown that the Higgs signal can still be observed well above the
background with the statistical error of the Higgs cross section at
the $10\div 30\%$ level in the wide range of Higgs mass $60\div
170$~GeV. The net error on $\Gamma(\hsm\to\gamgam)\br(\hsm\to b\anti
b)$ for $L=50\fbi$ is illustrated in Fig.~\ref{fig:gamma-gamma}.  
Thus, the error in the $\mhsm\lsim 120\gev$ mass
region will be in the $8\%\div 10\%$ range, rising to 15\% by
$\mhsm=140\gev$ and peaking at 30\% at $\mhsm=170\gev$, as illustrated
in Fig.~\ref{fig:gamma-gamma}.

For the Higgs bosons heavier that $2M_Z$ the Higgs signal in \GG\
collisions can be observed in $ZZ$ decay mode \cite{GunionHaber,BBC}
if one of the $Z$'s is required to decay to $l^+l^-$ to suppress the
huge tree-level $\gamma\gamma\to W^+W^-$ continuum
background. However, even though there is no tree-level $ZZ$ continuum
background, such a background due to the reaction $\gamma\gamma\to ZZ$
does arise at the one-loop level in the electroweak theory
\cite{AA-ZZ} which makes the Higgs observation in the $ZZ$ mode
impossible for $m_h\gsim (350\div 400)$~GeV. It was found that for
$185\lsim m_h\lsim 300$~GeV the $ZZ$ mode will provide a 10-20\%
determination of the quantity $\Gamma(h\to\gamma\gamma)\cdot BR(h\to
ZZ)$ (see Fig.~\ref{fig:gamma-gamma}).

The accuracies of the various measurements involved are a crucial
issue.  The errors for $\gamhsm$ are tabulated 
\footnote{For $\mhsm\gsim 130\gev$ one can compute
$\gamhsm=\Gamma(\hsm\to W\wstar)/\br(\hsm\to W\wstar)$ using LHC
data. Combined error for $\gamhsm$ is quoted in the
Table~\ref{errors} for this mass range.}
in Table~\ref{errors}.

\begin{table}[htb]
\caption[fake]{ The errors for $\Gamma(\hsm\to\gam\gam)$ as determined
using luminosity of $L=50\fbi$ accumulated in $\gam\gam$ collisions at
$\rts_{\epem}\sim \mhsm/0.8$ \cite{snow96}.  Approximate errors for
Higgs total width, branching ratios, and couplings-squared are given
for $L=200\fbi$ at $\protect\rts=500\gev$ NLC.  For
$\br(\hsm\to\gam\gam)$ the NLC and LHC results are combined.}

\footnotesize
\begin{center}
\begin{tabular}{|c|c|c|c|c|}
\hline
 Quantity & \multicolumn{4}{c|}{Errors} \\
\hline
\hline
{$\bf\mhsm$}{\bf (GeV)} & { \bf80} & { \bf100} & { \bf 110} & {\bf 120} \\
\hline
 $(\gam\gam \hsm)^2/(b\anti b\hsm)^2$ & $\pm 42\%$ & $\pm 27\%$ & $\pm 24\%$ &
 $\pm 22\%$ \\
\hline
 $\br(\hsm\to b\anti b)$ & \multicolumn{4}{c|}{$\pm5\%$} \\
\hline
 $\br(\hsm\to\gam\gam)$ & $\pm 15\%$ & $\pm 14\%$ & $\pm 13\%$ & $\pm 13\%$ \\
\hline
 $(\gam\gam\hsm)^2$ & \multicolumn{4}{c|}{$\sim \pm 12\%$}\\
\hline
 $\gamhsm$  & $\pm 19\%$ & $\pm 18\%$ & $\pm 18\%$ & 
 $\pm 18\%$ \\
\hline
 $(b\anti b\hsm)^2$ & $\pm 20\%$ & $\pm 19\%$ & $\pm 18\%$ & 
 $\pm 18\%$ \\
\hline
\hline
{$\bf\mhsm$}{\bf (GeV)} & {\bf 130} & {\bf 140} & {\bf 150} & {\bf 170} \\
\hline
 $(\gam\gam \hsm)^2/(b\anti b\hsm)^2$ & $\pm 23\%$ & $\pm 26\%$ & $\pm 35\%$ &
 $-$ \\
\hline
 $\br(\hsm\to b\anti b)$ & \multicolumn{2}{c|}{$\pm6\%$} & $\pm 9\%$ & $\sim
 20\%?$ \\
\hline
 $\br(\hsm\to\gam\gam)$ & $\pm 13\%$ & $\pm 18\%?$ & $\pm 35\%$ & $-$ \\
\hline
 $(\gam\gam\hsm)^2$ & $\pm 15\%$ & $\pm 17\%$ & $\pm 31\%$ & $-$ \\
\hline
 $\gamhsm$  & $\pm 13\%$ & $\pm 9\%$ & $\pm 10\%$ & 
 $\pm 11\%$ \\
\hline
 $(b\anti b\hsm)^2$ & $\pm 14\%$ & $\pm 11\%$ & $\pm 13\%$ & 
 $\pm 23\%$ \\
\hline
\hline
{$\bf\mhsm$}{\bf (GeV)} & {\bf 180} & {\bf 190} & {\bf 200} & {\bf 300} \\
\hline
 $(ZZ\hsm)^2$ & \multicolumn{2} {c|}{$\pm 4\%-\pm5\%$}& $\pm 6\%$ & $\pm 9\%$ \\
\hline
 $(\gam\gam\hsm)^2$ & $\pm 13\%$ & $\pm 12\%$ & $\pm 12\%$ &  $\pm 22\%$ \\
\hline
 $\gamhsm$  & $\pm 13\%$ & $\pm 14\%$ & $\pm 15\%$ & 
 $\pm 28\%$ \\
\hline
\end{tabular}
\end{center}
\label{errors}
\end{table}

\subsection{Measurements of the Higgs boson $CP$-properties}

The ability to control the polarizations of back-scattered photons
provides a powerful means for exploring the $CP$ properties of any
single neutral Higgs boson that can be produced with reasonable rate
at the Photon Linear Collider
\cite{CP}.
A $CP$-even Higgs bosons $h^0$, $H^0$ couple to the combination
$\vec{\varepsilon_1}\cdot \vec{\varepsilon_2}=-1/2
(1+\lambda_1\lambda_2)$, while a $CP$-odd Higgs boson $A^0$ couples to
$\vec{\varepsilon_1}\times \vec{\varepsilon_2}\cdot\vec{k_\gamma}=
\omega_\g/2 i\lambda_1 (1+\lambda_1\lambda_2)$, where
$\vec{\varepsilon_i}$ and $\lambda_i$ are photon polarization vectors
and helicities. The first of these structures couples to linearly
polarized photons with the maximal strength if the polarizations are
parallel, the letter if the polarizations are perpendicular.
Moreover, if the Higgs boson is a mixture of $CP$-even and $CP$-odd
states, as can occur {\it e.g.} in a general \thdm\ with $CP$-violating
neutral sector, the interference of these two terms gives rise to a
$CP$-violating asymmetries
\cite{CP}.  Two
$CP$-violating ratios could contribute to linear order with respect to
$CP$-violating couplings:
\be
{\cal A}_1=\frac{|{\cal M}_{++}|^2-|{\cal M}_{--}|^2}
                {|{\cal M}_{++}|^2+|{\cal M}_{--}|^2},
\quad
{\cal A}_2=\frac{2Im({\cal M}_{--}^*{\cal M}_{++})}
                {|{\cal M}_{++}|^2+|{\cal M}_{--}|^2}.
\ee
Since the event rate for Higgs boson production in $\g\g$ collisions
is given by
\begin{eqnarray}
dN &=& dL_{\g\g}dPS \frac{1}{4}
(|{\cal M}_{++}|^2+|{\cal M}_{--}|^2) \nonumber\\
&&\times
[(1+\langle\xi_2\tilde\xi_2\rangle)
+(\langle\xi_2\rangle+\langle\tilde\xi_2\rangle){\cal A}_1
+(\langle\xi_3\tilde\xi_1\rangle+\langle\xi_1\tilde\xi_3\rangle){\cal A}_2],
\end{eqnarray}
where $\xi_i$, $\tilde\xi_i$ are the Stokes polarization parameters,
two $CP$-violating asymmetries could be observed. The first one is
\begin{equation}
A_{circ} = \frac{N_{++}-N_{--}}{N_{++}+N_{--}}=
\frac{\langle\xi_2\rangle+\langle\tilde\xi_2\rangle}
{1+\langle\xi_2\tilde\xi_2\rangle}{\cal A}_1,
\end{equation}
where $N_{\pm\pm}$ correspond to the event rates for positive
(negative) initial photon helicities. Experimentally the measurement
of the asymmetry is achieved by simultaneously flipping the helicities
of both of the initiating laser beams.  Since the $A_{circ}$ is
proportional to the imaginary part of the SM contribution to the
$\g\g\to h^0$ amplitude, which is very small below $2\mw$ threshold,
this asymmetry can be useful only for $m_h\gsim 2\mz$.  
The asymmetry to be observed with linearly polarized photons is given
by
\begin{equation}
A_{lin} = \frac{N(\chi=\frac{\pi}{4})-N(\chi=-\frac{\pi}{4})}
{N(\chi=\frac{\pi}{4})+N(\chi=-\frac{\pi}{4})} = 
\frac{\langle\xi_3\tilde\xi_1\rangle+\langle\xi_1\tilde\xi_3\rangle}
{1+\langle\xi_2\tilde\xi_2\rangle}{\cal A}_2,
\end{equation}
$\chi$ is the angle between the linear polarization vectors of the
photons. The attainable degree of linear polarization $l_{\gamma}$ at
PLC depends on the value of $z_m=(\sqrt{s_{\g\g}})_{max}/2E_b$ which
can be changed in the case of free electron laser \cite{TEL97}. For
$z_m=0.82$ the degree of linear polarization is $l_\gamma\sim 0.33$
only, but $l_\gamma\gsim 0.8$ at $z_m\lsim 0.5$.  One finds \cite{CP}
that the asymmetries are typically larger than 10\% and are observable
for a large range of \thdm\ parameter space if $CP$ violation is present
in the Higgs potential.

\subsection{The discovery of the heavy states in extended Higgs models}

The PLC potential to discover Higgs bosons is especially attractive in
the search for heavy Higgs states in the extended models such as MSSM
\cite{GunionHaber,Gunion96}. The most important limitation of a
$e^+e^-$ collider in detecting the MSSM Higgs bosons is the fact that
they are produced only in pairs, $H^0A^0$ or $H^+H^-$ and the
parameter range for which the production process, $Z^*\to H^0A^0$ has
adequate event rate is limited by the machine energy to $m_{A^0}\sim
m_{H^0}\le \sqrt{s_{ee}}/2-20$~GeV ($m_{H^0}\sim m_{A^0}$ for large
$m_{A^0}$) \cite{Gunion96}. At $\sqrt{s_{ee}}=500$~GeV, this means
$m_{A^0}\le 230$~GeV. As $e^+e^-\to H^+H^-$ is also limited to
$m_{H^\pm}\sim m_{A^0}\le (220\div 230)$~GeV, it could happen that
only a rather SM-like $h^0$ is detected in $e^+e^-$ mode of the linear
collider, and none of the other Higgs bosons are observed. On the
other hand, $H^0$ and $A^0$ can be singly produced as $s$-channel
resonances in the \GG\ mode and PLC might allow the discovery of the
$H^0$ and/or $A^0$ up to higher masses
\cite{GunionHaber,Gunion96}. Particularly interesting decay channels
at moderate $\tan\beta$ and below $t\bar t$ threshold are $H^0\to
h^0h^0$ (leading to a final state containing four $b$ quarks) and
$A^0\to Zh^0$. These channels are virtually background free unless
$m_h^0\sim m_W$, in which case the large $\gamma\gamma\to W^+W^-$
continuum background would have to be eliminated by
$b$-tagging. Discovery of the $A^0$ or $H^0$ up to about
$0.8\sqrt{s_{ee}}$ would be possible. For large $\tan\beta$, the
detection of the $A^0$ or $H^0$ in the $b\bar b$ channel should be
possible for masses $\le 0.8\sqrt{s_{ee}}$
\cite{GunionHaber,Gunion96}, provided that effective luminosities as
high as 200~fb$^{-1}$ can be accumulated.

\begin{figure}[htb]
\setlength{\unitlength}{1in}
\begin{picture}(4.7,3.35)
\put(.8,0){\epsfig{file=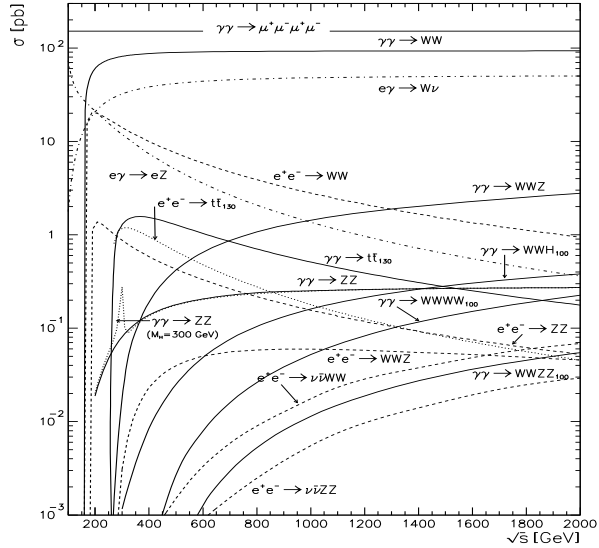,width=3.35in,height=3.35in}}
\end{picture}
\caption{ The cross sections of some processes in \ggam, $\gamma e$
and $\epem$ collisions.}
\label{fig:CS}
\end{figure}

\section{Gauge boson physics}

Without the discovery of a Higgs boson at LEP2, LHC or linear
collider, the best alternative to study the symmetry breaking sector
lies in the study of the self-couplings of the $W$.  The PLC will be
the dominant source of the $W^+W^-$ pairs at future linear colliders
due to the reaction \ggww\ with the large cross section, that fast
reaches at high energies its asymptotic value $\sigmaw =8\pi\alpha^2
/\mww \approx 81$~pb, which is at least an order of magnitude larger
than the cross section of $W^+W^-$ production in $e^+e^-$
collisions. With the rate of about 1--3 million of $W$ pairs per year
PLC can be really considered as a $W$ factory and an ideal place to
conduct precision tests on the anomalous triple
\cite{27,BAIL94,Trilinear} and quartic \cite{BAIL94} couplings
of the $W$ bosons.

The cross sections of main processes with the $W$ and $Z$ production
at PLC within SM are shown in Fig.~\ref{fig:CS} \cite{CS}.  When the
energy increases, the cross sections of a number of higher--order
processes become large enough.

In spite of enormous $WW$ event rates, prospects to improve the
precision of the measurement of the $W$ mass at LEP2 seem to be quite
limited. The reason is that the best estimated error on $\mw$ of
$30\div 40$~MeV \cite{MWLEP2} is extracted from the direct
reconstruction of invariant mass of the $W$ decay products by a
kinematic fit using the constraints of energy and momentum
conservation. Since the energy of colliding photons is not so
precisely fixed this method would not be effective at the PLC.

\begin{figure}[htb]
\setlength{\unitlength}{1in}
\begin{picture}(4.7,2.55)
\put(1.2,0){\epsfig{file=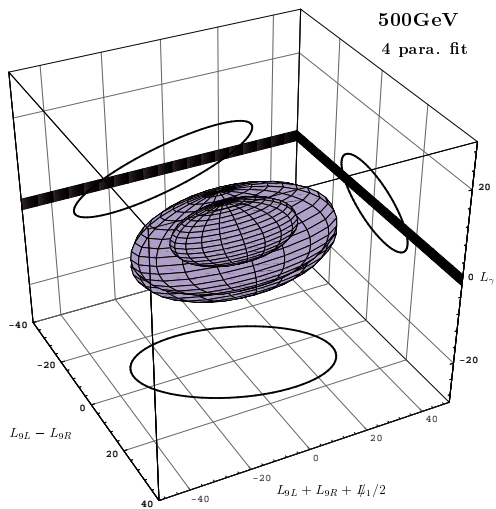,width=2.55in,height=2.55in}}
\end{picture}
\caption[fake]{The tri-dimensional bounds in the case of 4 different
anomalous couplings for a fit at $\rts=500$~GeV \cite{Trilinear}.  The
ellipsoid represents the $\epem$ constraints, while \protect\GG\
bounds are shown as two band projections on the planes.}
\label{fig:Ellipse}
\end{figure}

At $\rts=500$~GeV the benefits of the reaction $\g\g\to W^+W^-$ in precise
anomalous $W$ coupling measurements are
clearly visible \cite{Trilinear} in Fig.~\ref{fig:Ellipse}, since when
combined with the bounds from $\epem\to W^+W^-$ the parameter space
shrinks considerably. Since only one combination of triple anomalous
couplings (corresponding to $W$ anomalous magnetic moment) contributes
to the reaction $\g\g\to W^+W^-$ the allowed region is constrained to
be between two planes, while $\epem\to W^+W^-$ being sensitive to
several anomalous couplings constrains the parameter space outside
the ellipsoid.

With the natural order of magnitude on anomalous couplings
\cite{Fawzi-Talks}, one needs to know the SM cross sections with a
precision better than 1\% to extract these small numbers.  The
predictions for $W$ pair production, including full electroweak
radiative corrections in the SM are known with very little theoretical
uncertainty at least for energies below 1~TeV \cite{AA-WW}.

Although the cross section of $WW$ production is much larger in \GG\
collisions, this fact itself is not to be considered as an obvious
advantage of PLC.  The reason is that although the anomalous
contribution to the amplitude of longitudinal $W_LW_L$ pair production
is enhanced by a factor of $s/M_W^2$ both in \GG\ for $J_z=0$ and
$\epem$ collisions, the SM amplitude of $W_LW_L$ production at PLC is
suppressed as $M_W^2/s$, so that the contribution of the interference
term to the total cross section is decreasing as $1/s$ at PLC
\cite{Trilinear}. On the contrary, in $\epem$ collisions the anomalous
contribution is enhanced, corresponding to non-decreasing cross
section of $W_LW_L$ production. Recently the authors of
Ref. \cite{Trilinear} have demonstrated that enhanced coupling could
still be exploited in the \GG\ mode. Their clever idea is to
reconstruct the non diagonal elements of the $WW$ polarization density
matrix by analyzing the distributions of the decay products of the
$W$'s, thereby achieving the improvement over simple counting rate
method of more that an order of magnitude at $\rts=2$~TeV.  However,
although the benefits from \GG\ mode are evident at $\rts=500$~GeV
(Fig.~\ref{fig:Ellipse}), at energies above 1~TeV combining results
from $\epem$ and \GG\ modes does not considerably reduce the bounds
obtained from $\epem\to W^+W^-$ alone \cite{Trilinear}. This is
especially true for fits with one anomalous coupling. Qualitatively
these results can be understood considering the ratio $S/\sqrt{B}$ as
a measure of statistical significance of the anomalous coupling signal
$S$ with respect to the SM background $B$. Since the total SM cross
section is decreasing as $1/s$ in $\epem$ collisions and is constant
\GG\ collisions, while the enhanced anomalous cross section behaves
like a constant we get 
\be \frac{S(\epem\to W^+W^-)}{\sqrt{B(\epem\to W^+W^-)}} \propto \sqrt{s}, 
\ee 
while $S(\g\g\to
W^+W^-)/\sqrt{B(\g\g\to W^+W^-)} \propto 1$. If we take into account
that anomalous couplings affect mostly the cross section in the
central region, where the SM cross section behaves like
$\sigma(\g\g\to W^+W^-) \sim 8\pi \alpha^2/ p_T^2$, we get 
\be
\frac{S(\g\g\to W^+W^-)}{\sqrt{B(\g\g\to W^+W^-)}} \propto p_T, 
\ee
\ie\ the same improvement at higher energy as for $\epem$ collisions
but only for large values of $p_T$ cut $p_T\sim s$, with which the
cross section of $WW$ production in \GG\ collisions is not enhanced
any more with respect to production in $\epem$ collisions.

The process of $W$ production with the highest cross section in \GE\
collisions, \gewnu,\ with the asymptotic cross section of
$\sigma_{\g e\to W\nu} = \sigma_W /8\sin^2\theta_W\approx 43$~pb, 
is very sensitive to
the admixture of right--handed currents in W coupling with fermions
and could be also used to constrain the anomalous magnetic moment of $W$
\cite{27}.  Another example of the asymmetry, that could be used for
the measurement of the $W$-boson anomalous magnetic and quadrupole
moments has been proposed recently \cite{BrodskyRizzoSchmidt} and is
given by the so called polarization asymmetry
\be 
A^{+-}= \frac{\s_{++} -\s_{+-}}{\s_{++} +\s_{+-}}, 
\ee
where $\s_{\l_{\g}\l_{e}}$ is the polarized cross section of the
reaction $\g e\to W\nu$.  Using a quantum loop expansion it was shown
that there must be a center of mass energy where the polarization
asymmetry possesses a zero. The position of the zero may be determined
with sufficient precision to constrain the anomalous couplings of the
$W$ to better than the 1\% level at 500~GeV
\cite{BrodskyRizzoSchmidt}.  At higher energies the precise
measurements suffer from the same problems as those discussed for $W$
pair production in \GG\ collisions due to suppressed yield of $W_L$'s
with respect to $W_T$'s.

At higher energy the effective $W$ luminosity becomes substantial
enough to allow for the study of $W^+W^-\to W^+W^-$, $ZZ$
scattering in the reactions $\ggam\to WWWW$, $WWZZ$, when each
incoming photon turns into a virtual $WW$ pair, followed by the
scattering of one $W$ from each such pair to form $WW$ or $ZZ$
\cite{BAIL94,BRODHAW,JikiaWWWW}. The result is that a
signal of SM Higgs boson with $m_h$ up to 700~GeV (1~TeV) could be
probed in these processes at 1.5~TeV (2~TeV) PLC, assuming integrated
luminosity of 200~fb$^{-1}$ (300~fb$^{-1}$). However even larger
luminosity is needed in order to extract the signal of enhanced
$W_LW_L$ production in models of electroweak symmetry breaking without
Higgs boson \cite{JikiaWWWW}. The main problem is again
large background from transverse $W_TW_TW_TW_T$, $W_TW_TZ_TZ_T$
production.

\section{Conclusions}

Photon Linear Collider based on $\epem$ collider with $\rts=500$~GeV
\begin{itemize}
\item
provides unique opportunities to measure $\Gamma(h^0\to\g\g)$ up to
$m_h\lsim 350$~GeV, making possible with the use of NLC and LHC
measurements to measure $\Gamma_{tot}(h^0)$ and Higgs boson partial
widths;
\item
substantially extends NLC reach in discovering heavy Higgs states
$H^0$, $A^0$ in extended Higgs models such as MSSM or \thdm;
\item
can provide much more stringent bounds on $W$ anomalous
couplings complementary to those in $e^+e^-$ collisions.
\end{itemize}

One can fully exploit PLC potential at higher energies
$\sqrt{s_{\g\g}}\sim 1\div 2$~TeV if luminosity much higher than in
$\epem$ collisions is achievable \cite{TEL97}.

\section*{Acknowledgments}

I am grateful to the organizers of the Workshop for financial support
and warm hospitality. I would like to thank S.~Brodsky, T.~Han and
V.~Telnov for stimulating discussions. This work has been supported by
the Alexander von Humboldt Foundation.

\end{document}